\RequirePackage{fix-cm}
\documentclass[twocolumn]{svjour3}          % twocolumn
\smartqed  % flush right qed marks, e.g. at end of proof
\usepackage{graphicx}
\usepackage{caption}
\usepackage{subcaption}
\journalname{}
\begin{document}

\title{A turbulence model based on deep neural network considering the near-wall effect%\thanks{Grants or other notes
}
% \subtitle{A DNN-turbulence model considering the near-wall effect}


\author{Muyuan Liu         \and
        Yiren Yang         \and
        Hao Chen$^*$  %etc.\cormark[1]
}


\institute{$^*$Corresponding Author: Hao Chen \at
              School of Mechanics and Engineering, \\
              Southwest Jiaotong University, Chengdu 610031, China \\
              Tel.: +0086-28-87600797\\
%               Fax: +123-45-678910\\
              \email{chenhao@swjtu.edu.cn}           %  \\
%            \and
%            S. Author \at
%               second address
}

\date{Received: date / Accepted: date}

\maketitle

\begin{abstract}
There exists continuous demand of improved turbulence models for the closure of Reynolds Averaged Navier-Stokes (RANS) simulations. Machine Learning (ML) offers effective tools for establishing advanced empirical Reynolds stress closures on the basis of high fidelity simulation data. This paper presents a turbulence model based on the Deep Neural Network(DNN) which takes into account the non-linear relationship between the Reynolds stress anisotropy tensor and the local mean velocity gradient as well as the near-wall effect. The construction and the tuning of the DNN-turbulence model are detailed. We show that the DNN-turbulence model trained on data from direct numerical simulations yields an accurate prediction of the Reynolds stresses for plane channel flow. In particular, we propose including the local turbulence Reynolds number in the model input.
\keywords{turbulence modeling \and machine learning \and near-wall effect \and plane channel flow}
\end{abstract}

\section{Introduction}

\label{sec:intro}

Turbulent flows involve a range of spatial and temporal scales, a complete resolving of which is computationally expensive. Reynolds-averaged Navier-Stokes (RANS) simulation, which solves equations for mean quantities, is a feasible concept widely used in industrial turbulent flow problems. The Reynolds stresses are unknowns in the RANS equations and must be determined by a turbulence model. Linear eddy viscosity models (LEVM) assume a linear relationship between the Reynolds stress anisotropy tensor and the local mean strain rate:
\begin{equation}
\label{eq:1}
 a_{ij}=-2\nu_T\bar{\mathbf{S}},
\end{equation}
where $a_{ij}$ is the Reynolds anisotropy tensor with $a_{ij}=$ \newline $<u_i u_j>-\frac{2}{3} k  \delta_{ij}$ and $\bar{\mathbf{S}}$ the mean strain rate. The velocity co-variance $<u_i u_j>$ is the Reynolds stress tensor and $k$ the turbulent kinetic energy. Classical one-equation models (e.g. the turbulent-kinetic-energy model \cite{Kolmogorov1942} and the Spalart-Allmaras model \cite{spalart1994}) and two-equation models (e.g. the $k$-$\varepsilon$ model \cite{LAUNDER1974131} and the $k$-$\omega$ model \cite{wilcox1988}) based on the turbulent viscosity hypothesis mainly differ in the modeling of the turbulent viscosity $\nu_T$. The models with a linear stress-strain relationship do not capture the correct anisotropy of the Reynolds stresses in many flows including e.g. pipe flow with a contraction \cite{pope_2000}. 

Nonlinear turbulent viscosity models have also been developed for the closure problem of the RANS simulations, the general nondimensional form of which is given as \cite{pope_1975}:
\begin{equation}
\label{eq:2}
 b_{ij}=B_{ij}(\widehat{\mathbf{S}},\widehat{\mathbf{\Omega}}),
\end{equation}
where $b_{ij}$ is the Reynolds anisotropy tensor nondimensionalized by $k$. The mean strain rate and mean rotation rate tensors nondimensionalized by a turbulent time scale are denoted by $\widehat{\mathbf{S}}$ and $\widehat{\mathbf{\Omega}}$, respectively. The turbulent time scale can be constructed by means of the local turbulent dissipation rate $\varepsilon$ and the turbulent kinetic energy $k$, as suggested by Pope  \cite{pope_1975}. The explicit expressions of equation (\ref{eq:2}) have been proposed in a variety of different forms with examples found in \cite{pope_1975,gatski_speziale_1993,Rubinstein1990,CRAFT1996245}. Generally, the classical nonlinear viscosity models yield more accurate predictions of the Reynolds stress anisotropy and allow the calculation of secondary flows, yet are not widely used due to the inconsistent performance improvement  \cite{pope_2000,ling2016}.

It is well known that the turbulence modeling in the near-wall region should take the effect of the fluid viscosity into account, because the local turbulence Reynolds number $Re_{L}=k^2/(\varepsilon\nu)$ tends to zero approaching the wall, where $\nu$ denotes the kinematic viscosity of the fluid. Classically, the near-wall effect is accommodated by means of damping functions applied to the modeled isotropic turbulent viscosity $\nu_T$. As an example given by Jones and Launder \cite{JONES1972301}, in association with the $k$-$\varepsilon$ model the turbulent viscosity is given as
\begin{equation}
\label{eq:3}
 \nu_{T}=f_{\mu}C_{\mu}\frac{k^2}{\varepsilon},
\end{equation}
with a calibrated constant $C_{\mu}=0.09$ and a damping function
\begin{equation}
f_{\mu}= \mbox{exp}(\frac{-2.5}{1+Re_{L}/50}).
\end{equation}
Equation (\ref{eq:3}) reduces to the standard $k$-$\varepsilon$ formulation away from the wall. 
% In the past few years, the application of Machine Learning (ML) tools in the investigation of fluid flow problems has been a booming research field. A comprehensive review for this subject is found in \cite{Brunton2020}. Among other machine learning tools, Random Forest (RF)\cite{Wu2018,Wang2017}, Gene Expression Programming (GEP) \cite{Weatheritt2017} and Deep Neural Networks (DNN) \cite{Weatheritt2017,ling2016,Zhang2019} have been employed to construct nonlinear turbulence model that learns the stress-strain relationship from high fidelity simulation data, i.e. data from direct numerical simulation (DNS) or Large Eddy Simulation (LES). DNN has drew particular attention in the turbulence modeling because of the flexibility and simplicity of their model structure \cite{kutz_2017,Brunton2020}. 

Alongside with the growing popularity of applying machine learning methods in turbulence simulations (a recent review is given by Brunton et al. \cite{Brunton2020}), deep neural networks (DNN) have been introduced for developing RANS turbulent models in the past years. Deep neural network establishes a transformation of input features through multiple nonlinear interactions to an output, which enables the learning of nonlinear turbulence models from high fidelity simulation data, i.e. data from direct numerical simulations (DNS) or large eddy simulations (LES). Deep neural networks have gained attention in turbulence modeling partially due to its overwhelming performance in other research fields including e.g. image classification \cite{LeCun2015} and speech recognition \cite{Hinton2012}. Zhang and Duraisamy  \cite{ZhangAndDuraisamy2015} predicted a correction factor for the turbulent production term using neural networks. Ling et al. \cite{ling2016} designed a DNN architecture to model the turbulence closure which enables the reproduction of secondary flows in duct flow. Weatheritt et al. \cite{Weatheritt2017} applied DNN in turbulence modelling. Their model applied to jets in crossflow yields an improvement on the prediction of the Reynolds stress anisotropy over the model based on the linear relationship. Zhang et al. \cite{Zhang2019} predicted the Reynolds stress anisotropy in channel flows using DNN. 

Given an appropriate input, a nonlinear turbulence model based on DNN predicts the non-dimensionalized Reynolds-stress anisotropy tensor $b_{ij}$. One major concern of using DNN for turbulence modeling is the selection of the input features. In consistence with the classical non-linear turbulence models, almost all DNN-turbulence models select so far the local strain rate tensor $\widehat{\mathbf{S}}$ and rotation rate tensor $\widehat{\mathbf{\Omega}}$ as input features \cite{ling2016,Weatheritt2017,Zhang2019}. In addition to $\widehat{\mathbf{S}}$ and $\widehat{\mathbf{\Omega}}$, Zhang et al. \cite{Zhang2019} selected the wall units $y^+$ as an input feature, which takes into account the near-wall effect. Zhang et al. \cite{Zhang2019} showed that DNN with this additional input feature yields an improvement on the prediction of the Reynolds anisotropy tensor in plane channel flows. Alternatively, we propose to select the turbulence Reynolds number ($Re_{L}=k^2/(\varepsilon\nu)$) as an additional quantity in the input, in order to take into account the viscosity effect near the wall. We select the turbulence Reynolds number as the additional input feature, because the turbulence Reynolds number is a local quantity, which can be easily constructed with available turbulent kinetic energy $k$ and turbulent dissipation rate $\varepsilon$. Instead of $\widehat{\mathbf{S}}$ and $\widehat{\mathbf{\Omega}}$, equivalently, we select the nondimensionalized local velocity gradient for the input, because the strain rate and rotation rate tensors are a decomposition of the velocity gradient and using the velocity gradient reduces the number of items in the input. 

% We highlight that the PI-theory does permit this additional non-dimensional quantity based on the hypotheses of local dependence. 

% \subsection{Objective and organization}

The objective of this paper is to present a turbulence model based on DNN, which distinguishes from previous works mainly in the introduction of the local Reynolds number as an additional input feature. The DNN used in this paper is trained and tested on data obtained from direct numerical simulations of plane channel flows. The organization of this paper is as follows. We first detail the structure of the DNN for the turbulence modeling and the corresponding tuning process. Then, we evaluate the predicted Reynolds stress anisotropy given by the DNN-turbulence model followed by a conclusion.

\section{Deep Neural Network}
\label{sec:DNN}

\begin{figure}
 \centering
   \includegraphics[width=.7\linewidth]{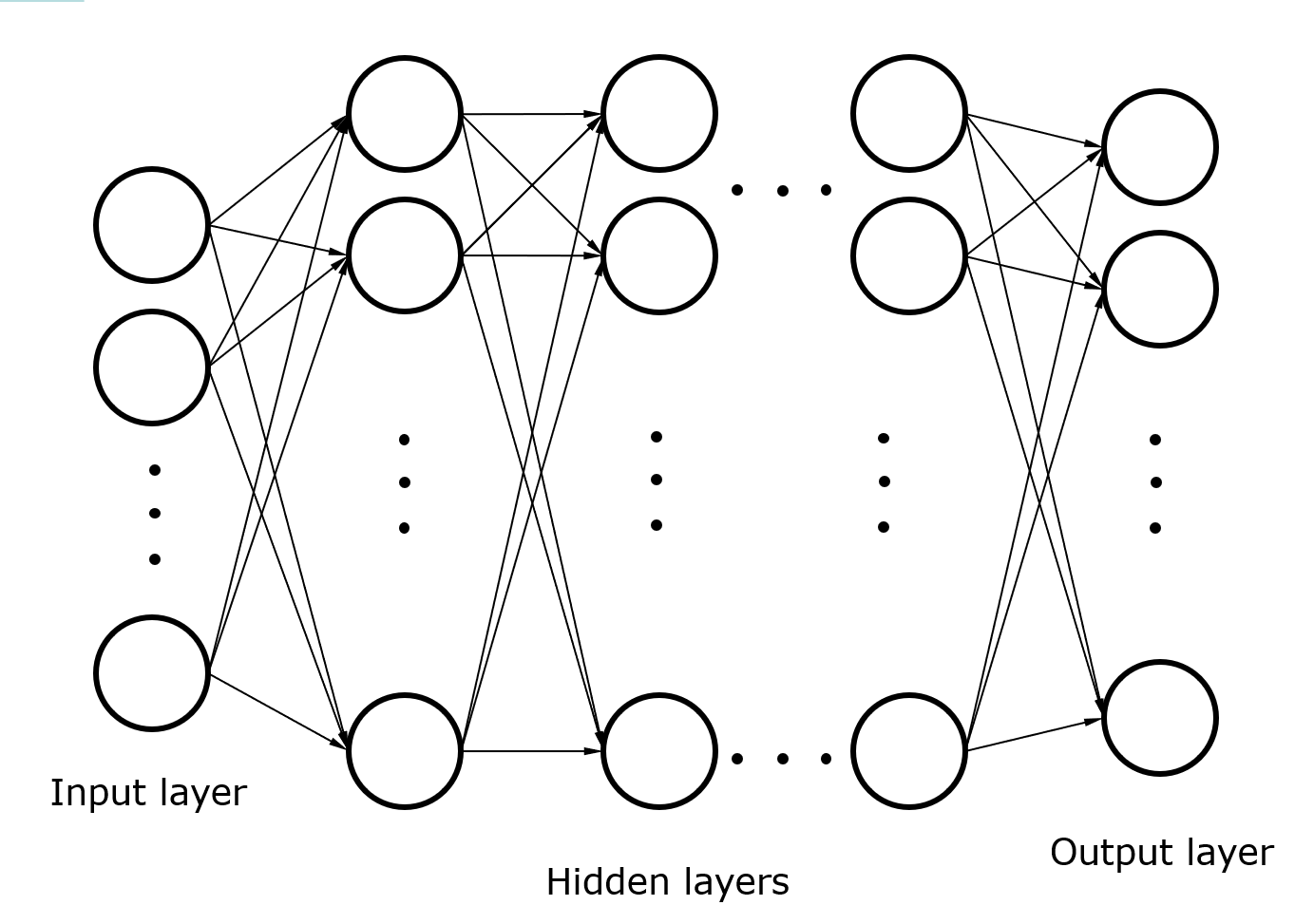}
  \caption{Structure of a deep neural network. }
  \label{fig:DNN}
\end{figure}

Deep neural networks are composed of multiple layers of nodes (or neurons), with each node connected to all nodes in neighboring layers, as shown schematically in Fig. \ref{fig:DNN}. The input layer at the far left is provided with an input $\mathbf{x}$ which is then linearly transformed by means of a weight matrix $\mathbf{W}$ represented by the connecting lines in Fig. \ref{fig:DNN} and a bias vector $\mathbf{b}$. The outcome $\mathbf{Wx}+\mathbf{b}$ is passed to the nodes in the first hidden layer, with each of the components then treated by means of a so-called activation function and serves as the input for the next layer. This procedure applies to the subsequent layers and terminates after giving an output vector $\mathbf{y}$ in the output layer. A common activation function $f=\mbox{max(0,x)}$, which is called Rectified Linear Unit (ReLU) \cite{maas2013rectifier}, is applied in this work. 

The objective of DNN is to learn a mapping $f: X\rightarrow Y$ on a training dataset constructed with values sampled from the input space $X$ and correspondingly from the output space $Y$. In order to propose a mapping, DNN minimizes a loss function in terms of $\mathbf{W}$ and $\mathbf{b}$ on the training dataset by means of gradient based methods. In this work, we use the mean squared error (MSE) between the predicted and sampled values as the loss function. The back propagation algorithm \cite{Rumelhart1986} calculates the gradients of the loss function in terms of the weights and the biases. The gradient decent algorithm then updates the weights and the biases by means of the obtained gradients multiplied by a learning rate $\eta$. We use the Adam method \cite{Kingma2014} which calculates an adaptive learning rate for the computation of the gradients.

% Adam was presented by Diederik Kingma from OpenAI and Jimmy Ba from the University of Toronto in their 2015 ICLR paper (poster) titled “Adam: A Method for Stochastic Optimization“.

\subsection{Data set}
\label{sec:dataset}

The dataset for training and testing the model is composed of statistical quantities obtained from direct numerical simulations of channels flows conducted by Lee and Moser and  \cite{lee_moser_2015} and Moser et al. \cite{Moser1999}, which are available under http://turbulence.ices.utexas.edu. Six flows characterized by Reynolds numbers computed on the basis of friction velocities are considered for assembling the dataset. The corresponding Reynolds numbers are $Re_{\tau}=390$, $550$, $590$, $1000$, $2000$ and $5200$ , respectively. The case with $Re_{\tau}=590$ is retrieved for composing the validation dataset, which serves for the determination of the hyper parameters of the DNN-turbulence model and the case with $Re_{\tau}=1000$ for the testing. The rest data consist of the training dataset.  

Based on the hypothesis of local dependency, one entry in the input for the DNN is composed of the local turbulence Reynolds number $Re_{L}$ and the local nondimensionalized velocity gradient, as argued in the introduction. The output is the Reynolds anisotropy tensor $b_{ij}$, where $b_{13}$ and $b_{23}$ are zero and excluded due to the symmetry of the geometrical configuration of plane channel flows. 

\subsection{Training of the DNN-turbulence model}

The training of a neural network is essentially an optimization process by means of updating the weights and biases. 
The initial values of the biases are not troublesome and simply set to zero, whereas an inappropriate initialization of the weights might lead to vanishing gradients that prevents the proceeding of the update \cite{Goodfellow-et-al-2016}. Following the suggestion given by He et al. \cite{He2015}, we use Gaussian distribution with zero means and standard deviations of the He-values for the initialization of the weights. The He-value is given by $\sqrt{2/n}$ with $n$ being the number of nodes in the precedent layer. This initialization method is generally applied in association with the ReLU activation function. 

Since training a neural network is analogous to fitting a regression model to the data, it suffers the risk of over-fitting, meaning obtaining high accuracy on the training data yet inaccurate on data not observed by the DNN in the training process. In order to suppress the over-fitting, we apply the weights decay \cite{Krogh1991ASW}, which adds a regularization term $\frac{1}{2}\lambda \mathbf{W}^T\mathbf{W}$ to the loss function.

Four hyper parameters are still to be determined, which are the coefficient $\lambda$ in the regularization term, the initial learning rate $\eta$, the number of hidden layers denoted by $nl$ and the number of nodes in each layer denoted by $nn$. In order to determine these hyper parameters, we sample the parameters randomly in a sample space around the parameters used by Zhang et al. \cite{Zhang2019} who trained a DNN for predicting the Reynolds anisotropy tensor in channel flows. Based on these samplings, we first select $\eta=2.5\cdot 10^{-7}$, which guarantees convergence and is not too small so that the computational time for a training is acceptable. Zhang et al. \cite{Zhang2019} observed that the Reynolds anisotropies predicted by a DNN-turbulence model yields unexpected oscillations due to over-fitting. We select $\lambda=0.001$ for the weight decay, which effectively reduces the over-fitting and thereby the oscillations.

With selected $\lambda$ and $\eta$, we generate a two dimensional grid with $nl=(3,4,5,6,7)$ and $nn=(10,20,30)$. We evaluate the root mean squared error (RMSE) loss on the validation dataset for each item on the grid. The RMSEs are calculated by summing over all non-zero and non-identical components of the symmetric tensor $b_{ij} $ in all entries of a dataset. Due to the random nature of training a neural network, we conduct the evaluation 10 times for each combination of $nl$ and $nn$. The averaged values are given in Table \ref{tab:grid comparison}. We select the trained DNN with $nl=5$ and $nn=30$ as the final DNN-turbulence model, which yields the smallest error on the validation dataset.  

\begin{table}
  \begin{center}
\def~{\hphantom{0}}  
    \begin{tabular}{rrrrrrrrr}
    \hline 
    
   $nl$                     & $nn=10$ & $20$   &$30$ &                       \\  
               \hline 
               
 $3$       & 0.0263 &  0.0234    & 0.0206                        \\      
 $4$    & 0.0252  &  0.0222    & 0.0202                       \\  
 $5$    & 0.0266   &  0.0204    & 0.0195                        \\   
 $6$    & 0.0235  &  0.0215    &  0.0198                     \\     
 $7$    & 0.0255 &  0.0209    & 0.0198                        \\
    \hline
    \end{tabular}
    \caption{Averaged RMSE evaluated on the validation dataset.}
    \label{tab:grid comparison} 
 \end{center}
\end{table}

\section{Prediction of the Reynolds anisotropy tensor }

\begin{figure}
\begin{subfigure}{0.5\textwidth}
  \centering
  \includegraphics[width=0.75\textwidth]{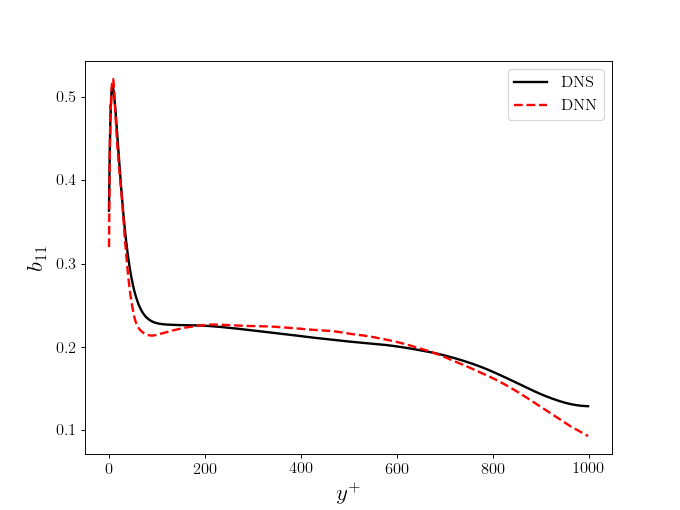}
%   \caption{}
  \label{fig:bij_1}
 \end{subfigure}%
 \newline
\begin{subfigure}{0.5\textwidth}
  \centering
  \includegraphics[width=0.75\linewidth]{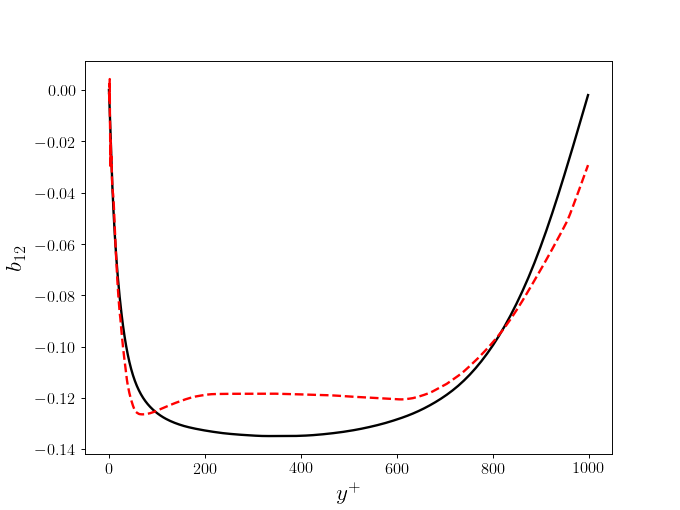}
%   \caption{}
  \label{fig:bij_2}
 \end{subfigure}
 \newline
\begin{subfigure}{0.5\textwidth}
  \centering
  \includegraphics[width=0.75\linewidth]{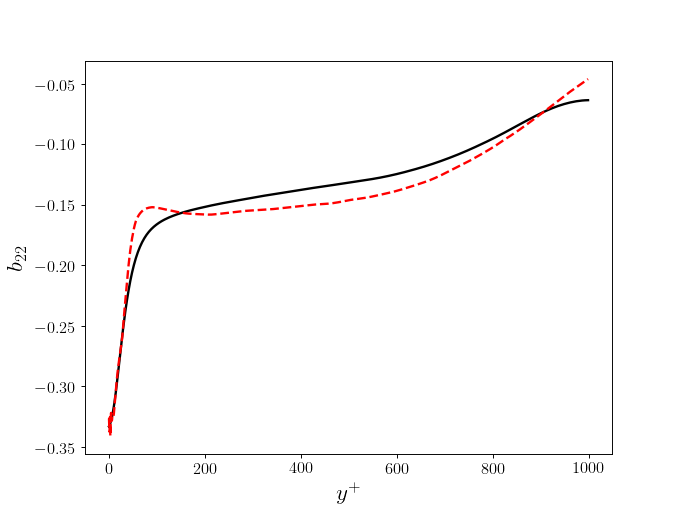}
%   \caption{}
  \label{fig:bij_3}
 \end{subfigure}%
 \newline
\begin{subfigure}{0.5\textwidth}
  \centering
  \includegraphics[width=0.75\linewidth]{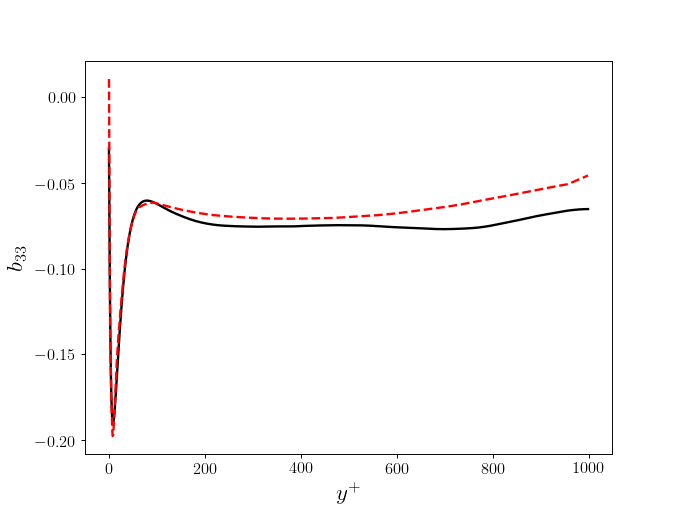}
%   \caption{}
  \label{fig:bij_4}
 \end{subfigure} 
 \caption{The Reynolds anisotropy tensor predicted by DNN and obtained from DNS.}
 \label{fig:bij}
\end{figure}

\begin{figure}
 \centering
  \includegraphics[width=0.75\linewidth]{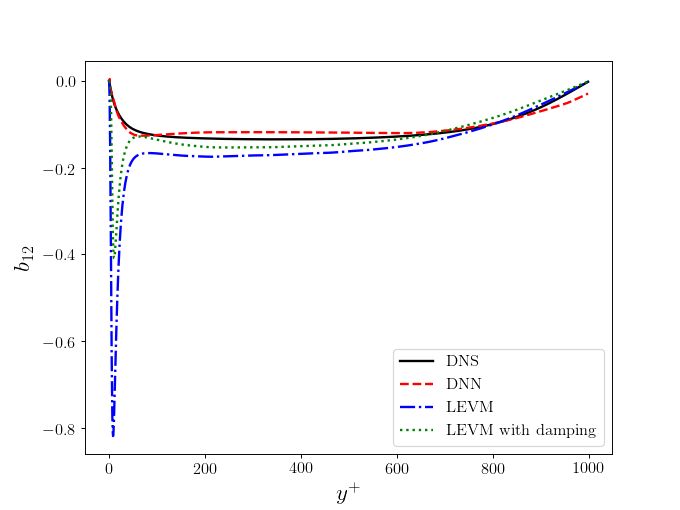}
  \caption{The non-zero component $b_{12}$ predicted by DNN, by linear model with and without damping compared to DNS values.}
  \label{fig:b12}
\end{figure}

The trained DNN represents a DNN-turbulence model which predicts the Reynolds anisotropy tensor given inputs that are not seen in the training process. In order to evaluate the prediction capability of this model, the predicted Reynolds anisotropy tensor is compared with their true values obtained from DNS for the test case $Re_{\tau}=1000$ in Fig. \ref{fig:bij}. Due to the symmetry of the geometry configuration of the plane channel flow as mentioned above, $b_{13}$ and $b_{23}$ have zero values and are not considered. It is shown that the DNN-turbulence model reproduces the Reynolds anisotropy tensor which is in very good agreement with the DNS-values. The classical linear models cannot predict the full anisotropy tensor and is restricted to predict $b_{12}$, because only $\widehat{S}_{12}$ in the strain rate tensor is not zero in plane channel flows. The components $b_{12}$ given by linear models with and without damping effect computed on the basis of equation (\ref{eq:2}) are plotted in Fig. \ref{fig:b12} along with the DNN prediction and DNS values. While the prediction given by the linear model without damping is acceptable in the inner region of the channel, the error in the near wall region is obviously large. The linear model with a damping yields better agreement in the whole channel, yet is clearly not as good as the prediction given by the DNN model especially in the near wall region. 
%  the only non-zero component of the strain rate tensor is $\widehat{S}_{12}$ in plane channel flows

A quantification of the prediction error in terms of the RMSE is given in Table \ref{tab:comparewithother}, along with RMSEs yielded by turbulence models based on Deep Neural Networks from previous works \cite{ling2016,Zhang2019}, though not of exactly the same geometrical and computational configuration. It is shown that the DNN applied in the present work achieves a significant improvement on the prediction accuracy, demonstrating the merit of using the turbulence Reynolds number as an additional input feature for data-driven turbulence modelling.
% closure problem.
% the DNN turbulence modeling.

\begin{table}
  \begin{center}
\def~{\hphantom{0}}  
    \begin{tabular}{llll}
    \hline 
  
   Flow &   RMSE \\
  \hline
  Duct flow \cite{ling2016}&  0.14   \\
  Flow over wavy wall \cite{ling2016}&   0.08 \\
%   Channel flow \cite{Zhang2019} &  & 0.06 \\
  Channel flow \cite{Zhang2019} (using $y^+$ ) & 0.05 \\
  Channel flow (This work) (using $Re_{L}$) & 0.02
  \\      
 
    \hline
    \end{tabular}
    \caption{Comparison of RMSEs of the Reynolds anisotropy tensor.}
    \label{tab:comparewithother} 
 \end{center} 
\end{table}
\newpage
\section{conclusion}

In conclusion, we trained a non-linear DNN-turbulence model which takes the near-wall effect into account by means of adding the local turbulence Reynolds number to the input feature. The model was shown to be able to predict the Reynolds stress anisotropy tensor accurately in a test channel flow. The proposed model has the potential to be deployed to similar and even more complicated flow problems, though the selection of the hyper parameters of the DNN and the model accuracy will be open questions. We will further address these issues next.

\section*{Acknowledgement}

This work was supported by the National Natural Science Foundation of China (Grant No. 11902275 and 11772273) and the Fundamental Research Funds for the Central Universities (Grant No. 2682020CX46).

%\begin{acknowledgements}
%\end{acknowledgements}

\section*{Conflict of interest}

The authors declare that they have no conflict of interest.

%  \bibliographystyle{unsrtnat}

% \bibliographystyle{spmpsci}      % mathematics and physical sciences
% \bibliographystyle{spphys}       % APS-like style for physics
% \begin{thebibliography}{}
% \bibliographystyle{unsrt}

% % Format for Journal Reference
% Author, Article title, Journal, Volume, page numbers (year)
% Author, Book title, page numbers. Publisher, place (year)
% etc
% \bibliography{cas-refs}{}
% \bibliographystyle{cas-model2-names}
% \end{thebibliography}

\end{document}